 \documentclass[pmlr,twocolumn,10pt]{jmlr} 

\usepackage{booktabs}
\usepackage{rotating}

\usepackage[load-configurations=version-1]{siunitx} 

\usepackage[utf8]{inputenc} 
\usepackage[T1]{fontenc}    
\usepackage{hyperref}       
\usepackage{booktabs}       
\usepackage{amsfonts}       
\usepackage{nicefrac}       
\usepackage{microtype}      
\usepackage{comment}
\usepackage{floatrow}   
\usepackage[framemethod=tikz]{mdframed}
\usepackage{graphics}
\usepackage{adjustbox}
\usepackage{tikz} 
\usepackage{placeins}

\graphicspath{ {./figs/} }

\title{Stereo Video Reconstruction Without Explicit Depth Maps for Endoscopic Surgery}

\author{%
    \Name{Annika Brundyn} \Email{annika.brundyn@nyu.edu} \\
    \Name{Jesse Swanson} \Email{jesse.swanson@nyu.edu} \\
    \Name{Kyunghyun Cho} \Email{kyunghyun.cho@nyu.edu} \\
    \addr Center for Data Science, New York University
    \AND
    \Name{Doug Kondziolka} \Email{douglas.kondziolka@nyulangone.org}\\
    \Name{Eric Oermann} \Email{eric.oermann@nyulangone.org} \\
    \addr Department of Neurosurgery, Langone Health, New York University
}

\begin{document}

\maketitle

\begin{abstract}
We introduce the task of stereo video reconstruction or, equivalently, 2D-to-3D video conversion for minimally invasive surgical video. 
We design and implement a series of end-to-end U-Net–based solutions for this task by varying the input (single frame vs. multiple consecutive frames), loss function (MSE, MAE, or perceptual losses), and network architecture. 
We evaluate these solutions by surveying ten experts – surgeons who routinely perform endoscopic surgery. 
We run two separate reader studies: one evaluating individual frames and the other evaluating fully reconstructed 3D video played on a VR headset.
In the first reader study, 
a variant of the U-Net that takes as input multiple consecutive video frames and outputs the missing view performs best. 
We draw two conclusions from this outcome. First, motion information coming from multiple past frames is crucial in recreating stereo vision. Second, the proposed U-Net variant can indeed exploit such motion information for solving this task.
The result from the second study further confirms the effectiveness of the proposed U-Net variant. The surgeons reported that they could successfully perceive depth from the reconstructed 3D video clips. They also expressed a clear preference for the reconstructed 3D video over the original 2D video. 
These two reader studies strongly support the usefulness of the proposed task of stereo reconstruction for minimally invasive surgical video and indicate that deep learning is a promising approach to this task.
Finally, we identify two automatic metrics, LPIPS and DISTS, that are strongly correlated with expert judgement and that could serve as proxies for the latter in future studies.
\end{abstract}

\section{Introduction}
\label{sec:intro}
\begin{figure}[htbp]
\floatconts
    {fig:stereo}
    {\caption{Proposed task of stereo reconstruction to facilitate depth perception in endoscopic surgery. Adapted from \citet{endo_figure} and \citet{stereo_figure}.}}
    {\includegraphics[width=\linewidth]{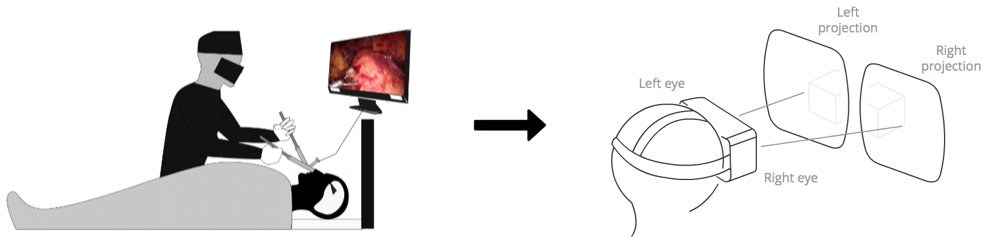}}
\end{figure}

The development and use of endoscopes for minimally invasive surgery was one of the major surgical innovations of the 20th century \citep{Li2005-lf}. An endoscope is a thin, tubular instrument with an attached camera. During surgery, an endoscope is inserted through a small incision or natural orifice so that a surgeon is able to view the internal area and operate on it by way of tiny surgical instruments (see \figureref{fig:stereo}). This obviates the need for large incisions, reducing patient healing time and the risk of post-operative infection. Technological advances in optics, illumination, and miniaturization over the past hundred years have expanded the reach of endoscopy, which has in turn revolutionized surgery \citep{berci2000_history}.

A major drawback of early endoscopy was the lack of depth perception arising from 2D vision. Although many surgeries can safely be performed using a 2D endoscope, it is more challenging to operate around critical and highly complex physiological structures in this setting. For example, it is difficult for a surgeon to judge distances between a brain tumor and the surrounding nerves and tissue without depth perception.

Such difficulty is naturally resolved with 3D stereo imaging. Furthermore, 3D stereo endoscopes have a shallower learning curve and can help novice surgeons take advantage of minimally invasive techniques \citep{Koc2006-lc}. While modern 3D stereo endoscopes exist, they cannot be miniaturized to the same degree as 2D endoscopes (4mm vs 2.7mm in diameter for the smallest in each class). They also suffer from a smaller field of view \citep{fov}, rendering them infeasible for some types of surgeries.

In this paper, we frame the task of automatically converting 2D endoscopic video to a 3D format as stereo video reconstruction.
3D content is typically stored in stereo format: i.e., two perspectives of the same scene. When viewed together, the disparity between the two perspectives simulates natural binocular vision, resulting in a 3D experience \citep{xie2016deep3d}. In the absence of a 3D stereo endoscope with two cameras, one can generate an alternate view of the existing 2D input, which, when combined with the original view, forms a stereo pair. The result can be viewed using 3D glasses or a head-mounted VR display (see \figureref{fig:stereo}), effectively recreating 3D depth perception for the operating surgeon.

We propose a deep neural network (DNN) as a solution to this task. In our approach, a DNN is trained to reconstruct an alternative perspective of the current frame using the information available in past video frames. This is necessary, as we will demonstrate empirically, since creating an alternative perspective from a single view is a highly under-constrained problem. Past frames may contain information about occluded regions in the current view and may capture more accurate depth information.

We run a series of experiments with DNNs to determine whether multiple past frames improve 3D stereo reconstruction and to identify an optimal network architecture and optimal learning parameters. We conduct an extensive and rigorous evaluation of these DNN variants through two sets of reader studies involving experienced surgeons. Our reader studies confirm that DNNs designed to leverage temporal information from past frames result in superior stereo reconstruction when compared to a single-frame approach. We also test a diverse set of automated metrics and correlate them against the outcome of these reader studies. We identify two perceptual metrics, DISTS \citep{Ding2020DISTS} and LPIPS \citep{zhang2018lpips}, that correlate strongly with expert judgement. 

The task and the approach presented in this paper are not restricted to endoscopic video but rather generally applicable to any video content. We find this particular use case interesting and technically meaningful because factors such as sharpness, hallucinations, and accurate depth perception are more salient here than in many other applications. We anticipate, however, that the findings from this work will be applicable and transferable to other similar use cases in the future.

\section{Related Work}

Here, we review related prior work in 2D-to-3D reconstruction, applications in endoscopy, and evaluation metrics for the task.

\subsection{2D-to-3D Reconstruction}

Traditional 2D-to-3D reconstruction methods often consist of two stages. First, a depth map is constructed from the 2D input; then, a depth image-based rendering (DIBR) algorithm combines the depth map with the input view to generate the missing view of the stereo pair \citep{Fehn}. Depth maps can be constructed using various techniques, among them manual construction by artists, structure-from-motion (SfM) approaches, and, more recently, machine learning algorithms \citep{Li_depth_maps, mei2002, Ummenhofer2016}. 

Monocular (i.e., 2D) depth estimation is itself a difficult task. Recent trends in deep learning have shifted toward training end-to-end differentiable systems, allowing explicit depth estimation to be bypassed. Accordingly, our proposed model only requires stereo pairs for training and learns to directly predict the right view from the left view. This is similar to Deep3D \citep{xie2016deep3d}, a stereo reconstruction model trained without supervisory ground-truth depth maps. Deep3D uses a single RGB frame as input, but the authors suggest incorporating temporal information from multiple frames as a promising direction for future research. In our work, we find that using multiple frames as input to our model achieves superior quantitative and qualitative performance when compared to single-frame models. 

Our method is built as a fully convolutional encoder–decoder architecture, largely adapting the U-Net architecture \citep{ronneberger2015unet}. U-Nets have been used in depth estimation \citep{wiles2020} and novel-view synthesis \citep{kim2020novel}. They have also been applied to the task of semi-supervised video segmentation \citep{sibechi2019exploiting}.

\subsection{Deep Learning Applications to Endoscopy}

Most of the existing work related to this paper focuses on depth estimation and topographical reconstruction in endoscopy \citep{MAHMOOD2018230,10.1117/12.2293785,10.1007/978-3-030-01201-4_15,turan2018sparse,lui2019}. For instance, \citet{KUMAR2014862} design a multi-stage procedure – involving 3D shape reconstruction, registration with a 3D CT model, endoscope position tracking, and depth map calculation – to allow for stereo image synthesis in laparoscopic surgery. These stages are often developed and tuned separately from one another and require expensive ground-truth annotations. Unlike these earlier studies, we investigate an end-to-end approach that requires minimal engineering and annotation.

It is tempting to generalize the success of deep learning for ``natural'' video to surgical video. It is, however, unclear whether this generalization is reasonable, since there are many features that are specific to surgical video that are not present in natural video. These include soft textures, homogeneous colors, inconsistent sharpness, variable depth of field and optical zoom, inconsistent motion, and obstructions arising from fluid and smoke \citep{Petscharnig2018}. Therefore, it is necessary to investigate the applicability and effectiveness of deep learning–based end-to-end approaches to 2D-to-3D reconstruction in surgical video, as we do in this paper.

\subsection{Evaluation Metrics}

Quantitatively assessing the quality of generated stereo video an open problem. 
In our own evaluation, we simplify the problem by assessing the image quality of the generated frames rather than the full video, following the convention of previous work (see \citet{xie2016deep3d}). 
Various automated metrics have been proposed to assess the quality of generated images, including 
structural similarity index (SSIM), peak signal-to-noise ratio (PSNR), visual information fidelity (VIF), and Fr\'echet inception distance (FID) \citep{Li_2019_CVPR_Workshops, song2014}. More recently, a family of metrics called \textit{perceptual} metrics have been found to closely correspond with human qualitative assessments \citep{ding2021_iqa}. 
In this first attempt at establishing the suitability of automated metrics in the domain of endoscopic surgery, we correlate a diverse set of automated metrics against human-perceived quality assessments obtained via a series of reader studies with ten experienced surgeons.

\section{Methods}
In this section, we describe the task of generating stereo video from 2D video and present a neural network–based approach devised for this purpose.

\subsection{Task Description}

We refer to ``stereo vision'' as a pair of images (or video frames), corresponding to the left and right views of the same scene, and define the task of 2D-to-3D reconstruction, or equivalently stereo video reconstruction, as the problem of generating a missing view (either left or right) given the other (observed) view. 
Without loss of generality, we take the left view as the observed input and the right view as the missing view to be reconstructed. In the case of video, we have multiple past frames of the left view available, while we do not assume the availability of any right-view frames. 

More specifically, 
we propose a solution to the task of {\it online} {\it frame-level} stereo video reconstruction. 
Given a sequence of left-view frames $\mathbf{x}_{\leq t}$ up to time $t$, we predict the corresponding right view $\mathbf{y}_t$ at time $t$. 

Surgical video can be of arbitrary length, and it is unlikely for distant past frames to be relevant to the current scene. We therefore consider only a few frames $K$ from the immediate past as the input. In other words, a system solving this task takes the most recent $K$ frames, including the current frame, $\mathbf{x}_{t-K+1:t} \in \mathbb{R}^{K \times 3 \times H \times W}$, and outputs the right view $\mathbf{y}_t \in \mathbb{R}^{3 \times H \times W}$, where we assume three-channel RGB frames of width $W$ and height $H$.

\begin{figure}[htbp]
\floatconts
  {fig:train}
  {\caption{Proposed approach. Models take as input a variable number of consecutive video frames from the left view and directly predict the missing right view for the current (latest) frame.}}
  {\includegraphics[width=\linewidth,trim={0cm 2.5cm 0cm 0cm}]{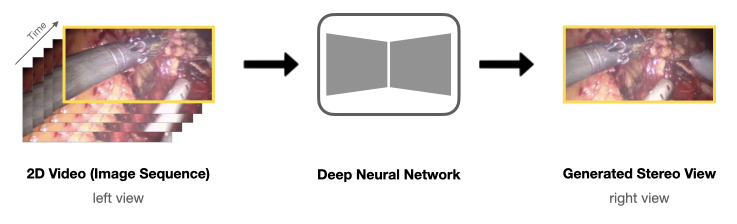}}
\end{figure}

\subsection{Deep Learning for 2D-to-3D Video Reconstruction}

We present a DNN that takes as its input the left view and directly renders the right view. 
We train this neural network 
using a dataset of 3D endoscopy procedures, to 
minimize the reconstruction error between the generated and the true right view. 
We do not require any explicit depth information or intrinsic camera parameters.

\subsubsection{A Base Network Architecture: A Modified U-Net}

We modify a fully convolutional U-Net as the base architecture for the task of stereo view generation \citep{ronneberger2015unet}. Our U-Net implementation differs from the original architecture in two ways: 1) decoder blocks use padding instead of cropping to concatenate features from the skip connections, and 2) the decoder uses either strided transpose convolution or bilinear interpolation for upsampling. We denote this U-Net model as $g_\theta$. 
As in the original implementation, we do not include any batch normalization layers. 
The 2D-to-3D reconstruction task is, then, to predict the corresponding right view given the left-view frames: i.e., $\mathbf{y}_t = g_{\theta}\left(\mathbf{x}_{t-K+1:t}\right)$.

\subsubsection{Capturing Temporal Information}

Unlike conventional applications of the U-Net, such as semantic segmentation \citep{ronneberger2015unet, siddique2020unet}, our modified U-Net must handle a temporal series of left-view frames. The temporal structure contained in the input frames can be exploited by the U-Net in different ways, and we explore three possible approaches in this paper.

\paragraph{(1) Current Frame Only.}

The simplest, but least effective, approach is to discard all the past frames and keep only the current frame as input ($K = 1$) to the network $g_\theta$, in order to predict the corresponding stereo frame $\mathbf{y}_t$. We use this approach to verify our hypothesis that temporal information from the observed view is useful for reconstructing the missing view, which has not yet been confirmed or refuted in the context of surgical video.

\paragraph{(2) Stacking Multiple Frames.}

We present the temporal ordering of observed frames by stacking them along the channel axis, similarly to color. In other words, we reshape the 4D input tensor $\mathbf{x}_{\leq t} \in \mathbb{R}^{K \times 3 \times H \times W}$ into a 3D tensor of size $3K \times H \times W$ and feed this 3D tensor to the U-Net architecture $g_\theta$, as if it were an image with $3 \times K$ color channels.

\paragraph{(3) A Spatio-Temporal U-Net.}

Like spatial structures, temporal structures exhibit themselves at multiple scales. Pixel-level temporal information allow fine-grained textures to be inferred, while object-level temporal information can help overcome occlusion in individual frames. In order to better incorporate the multi-scale nature of temporal information, we modify the U-Net architecture by inserting a ``temporal module'' at each scale between the encoder and decoder, capturing unique aspects of temporal information at each scale. 

More specifically, each of the $K$ consecutive frames in the input $\mathbf{x}_{t-K+1:t}$ is fed to the encoder separately and results in $K$ spatial feature maps. This sequence of feature maps is processed by the temporal module, before being sent to the corresponding layer at the decoder for concatenation. For a graphical illustration of the method, see can be seen in \figureref{fig:spatiotemp_unet}. In this work, we use a single 3D convolutional layer as the default temporal module. We compare this module against two naive approaches: the element-wise average and element-wise maximum modules.

\begin{figure}[htbp]
\floatconts
  {fig:spatiotemp_unet}
  {\caption{Spatio-temporal U-Net architecture.}}
  {\includegraphics[width=\linewidth,trim={1.5cm 2cm 0cm 0cm}]{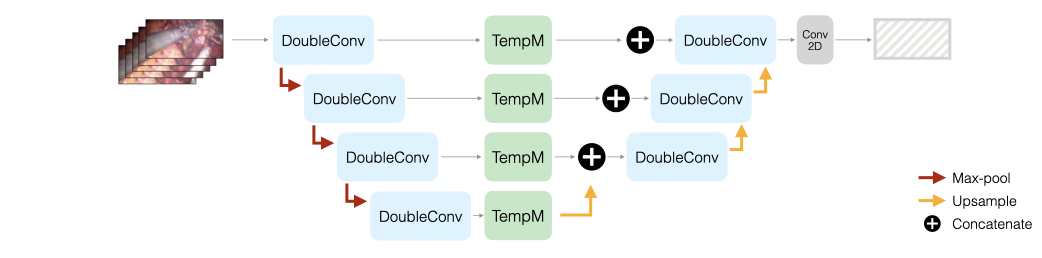}}
\end{figure}

\subsection{Loss Functions} 

The choice of loss function is known to have a large impact on the quality of generated images. Pixel-wise loss functions, such as mean squared error (MSE) and mean absolute error (MAE),
correlate poorly with perceived image quality \citep{ding2021_iqa}. However, when these losses are computed in perceptually appropriate representation space, they have been found to correlate better with human perception. 

In our experiments, we investigate three loss functions: MSE, MAE, and a perceptual loss function. We use the sum of the MAEs computed using the feature maps extracted from the first three blocks of VGG16 \citep{simonyan2015deepvgg}, pretrained on the ImageNet dataset \citep{deng09_imagenet} for the perceptual loss.

\section{Experimental Setup}

In this section, we describe the training procedure for our DNN-based computational models, and the subjective testing procedure used to collect human ratings of the generated images and VR videos. We ensure the reproducibility of our experiments by reporting on public datasets, and by publicly releasing our code.

\subsection{Dataset}

We use the {\it da Vinci} endoscopic dataset from the Hamlyn Center for Robotic Surgery \citep{ye2017selfsupervised}. This dataset consists of rectified stereo images of size $384 \times 192$, partitioned into 34,241 train and 7,192 test image sequences from video. Each video was captured \emph{in vivo} during a partial nephrectomy procedure performed using a da Vinci surgical system. Details regarding the how the specific train and validation dataset splits were obtained is discussed in \appendixref{data}.

\subsection{Training}

To investigate the effect of using multiple frames for 2D-to-3D reconstruction, we try different training set-ups. Specifically, we vary the number of input frames between one (current only), five, and ten consecutive frames. We also test five- and six-layer U-Nets, to understand whether more layers allow better temporal information to be captured.

All networks are trained with 
the Adam optimizer~\citep{kingma2017adam} 
from scratch, with a constant learning rate of $0.0001$. Each model is trained on a single NVIDIA V100 GPU for no more than 30 hours. 
We used a mini-batch size of 16, although we sometimes use smaller mini-batches to work with the limited onboard memory (e.g., when we train a network with the 10-frame input). For memory efficiency, we use 16-bit precision wherever possible.

\subsection{Constant Shift Baselines}
\label{sec:baseline_setup}
In order to ensure that the proposed U-Net–based approach performs non-trivial reconstruction, we include two baselines for comparison. 
In both cases, we shift all the pixels in the current left-view frame by a constant disparity $\delta$. The disparity $\delta$ is estimated by minimizing the loss (MSE, MAE, or perceptual) on the validation set. These two baselines differ in that one fills the newly added pixels with zeros (i.e., the color black), while the other fills the missing pixels with the original values from the input, resulting in a duplicated strip on one side of the generated frame.

\subsection{Automatic Evaluation Metrics}

Image-quality assessment is an active area of research \cite{ding2021_iqa}. Widely used evaluation protocols for video generation often rely on image-similarity-based metrics \cite{oprea2020review}. We select and report five image-quality metrics – PSNR, SSIM~\citep{wang2004ssim}, FID~\citep{heusel2018fid}, LPIPS~\citep{zhang2018lpips}, and DISTS~\citep{Ding2020DISTS} – for the purpose of understanding these metrics' correlation with human perception in the context of surgical endoscopy video.

\subsection{Expert Evaluation: Reader Studies}

We conduct two main reader studies. In the first reader study, we focus on evaluating individual predictions from the U-Net variants. Each expert is asked to choose between two candidate frames, based on their quality. In the second study, we model a more realistic scenario in which we ask experts to assess the quality of stereo-vision video using a VR kit. Due to time constraints, we use a subset of promising models from the first reader study for 3D-video-quality assessment.

\subsubsection{Frame-Level Reader Study}

\paragraph{Selected Models.}

We select 
eight models for the frame-level reader study. 
Our selection is based on a mix of automatic evaluation metrics and qualitative feedback from a supervising surgeon. Descriptions of the eight models, as well as further detail regarding the selection process, can be found in \appendixref{sec:app_screening}. 
The only multi-frame models selected were those that used the spatio-temporal U-Net architecture with a single 3D convolutional layer as the temporal module. Initial evaluation showed that stacking multiple frames resulted in similar or worse performance compared to using just a single frame. Spatio-temporal networks with an average or maximum temporal module produced visible inconsistencies such as double vision and were thus not included in the reader study.

\paragraph{Reader Study Design.}

To acquire human perceptual assessment of the generated results at the frame level, we conduct a reader study with 10 experts. Experts are surgeons with between 6 and 30 years of medical experience, and all have experience performing endoscopic surgery. 

We use the two-alternative forced choice (2AFC) method to compare the eight different models. 
Each model is compared against every other model three times, using the same three examples. Additionally, each model is compared against the target once, resulting in a total of 92 comparisons.
In a given trial, a surgeon is shown two frames generated by two distinct models (candidate synthetic right-view frames), as well as the corresponding left view that was used as the input to the models. 
The surgeon is then asked the following question: \textit{Given the left view, which of the generated right views (A or B) is of better quality?} 
Each image is shown in the native resolution of the dataset ($384\times 192$ pixels), with the option to zoom in on any portion of the presented frame. 
No time limit is applied.
An example is shown in \figureref{fig:image_survey} in \appendixref{sec:image_survey_more}. We employ the Bradley-Terry model \citep{bradley1952} to convert pair-wise comparison results to global rankings and calculate log-worth scores. Further details can be found in \appendixref{sec:bradleyterry}.

\subsubsection{Video-Level Reader Study: VR}

The quality of individual frames is an insufficient proxy for stereo video quality. Successful stereo video reconstruction requires that the viewer's perception of 3D depth remain consistent across consecutive frames. The shift between the two views that form a stereo pair is subtle. It is difficult to assess from a single left view whether a given right view would in fact result in realistic 3D depth perception when viewed in stereo. Moreover, presented with only a single stereo pair, it is impossible to judge whether the model achieves consistent reconstruction across time. 

We select five models from the first reader study and further evaluate them using a VR kit. Experts are first presented with a 25-second 2D clip. They then watch the same clip in 3D, generated from one of the selected models. The 3D videos are played using a Google Cardboard VR headset in conjunction with an iPhone 12. Experts are asked 1) whether the 3D video provided a better viewing experience than the 2D video did, and 2) to rate the quality of the reconstructed 3D video on a scale from 1 (worst) to 5 (best).

\section{Results and Discussion}

\begin{table*}[htbp]
\floatconts
  {tab:models}
  {\caption{Comparison of model architectures, ranked according to reader study results. We calculate the percentage of times (out of 24) a given model wins the majority vote and report this in the column "win \%". The top two models as ranked by each metric are indicated in bold.}}
  {\begin{tabular}{p{0.6cm}lrrrrrrr}
    \toprule
    & & \multicolumn{2}{c}{Reader Study} & \multicolumn{5}{c}{Automated Metrics}\\
    \cmidrule(r){3-4}
    \cmidrule(r){5-9}
    Rank & Model & log-worth & win \% & DISTS $\downarrow$ & LPIPS $\downarrow$ & FID $\downarrow$ & PSNR $\uparrow$ & SSIM $\uparrow$  \\
    \midrule
    1 & 5 fr+perceptual+6 layers    & 0.00  $\pm$ 0.20 & 100 & \textbf{0.110} & \textbf{0.116}& 50.53         & 22.88         & 0.627         \\
    2 & 5 fr+perceptual             & -0.91 $\pm$ 0.17 & 86  & \textbf{0.116} & \textbf{0.119}& \textbf{48.42}& 22.77         & 0.616        \\
    3 & 5 fr+perceptual+bilinear    & -1.85 $\pm$ 0.16 & 52  & 0.117          & 0.124         & 53.68         & \textbf{23.25}& 0.710         \\
    4 & 1 fr+perceptual             & -2.04 $\pm$ 0.16 & 52  & 0.119          & 0.125         & 52.56         & 22.63         & 0.624         \\
    5 & 10 fr+perceptual            & -2.41 $\pm$ 0.16 & 38  & 0.120          & 0.131         & \textbf{50.16}& 22.63         & 0.620         \\
    6 & 1 fr+MSE+6 layers           & -2.55 $\pm$ 0.16 & 48  & 0.143          & 0.156         & 65.78         & 23.04         & 0.700         \\
    7 & 1 fr+MAE                    & -3.74 $\pm$ 0.20 & 19  & 0.140          & 0.156         & 74.66         & 23.14         & \textbf{0.716}\\
    8 & 5 fr+MSE+6 layers           & -4.42 $\pm$ 0.23 & 5   & 0.143          & 0.159         & 66.28         & \textbf{23.78}& \textbf{0.722}\\
    \bottomrule
  \end{tabular}}
\end{table*}

\subsection{Frame-Level Reader Study}

\paragraph{Multiple Frames and Perceptual Loss.}

A major finding from the first reader study is that the proposed U-Net variant benefits from the availability of multiple left-view frames as input. This is clear from the top-three entries in \tableref{tab:models}, according to both the log-worth and the win rate. All three models take five left-view frames as input. This observation confirms our earlier hypotheses that 1) multiple frames from one view facilitate reconstructing the other view, and 2) the proposed variant of the U-Net is capable of exploiting such temporal information from multiple frames.

Another finding – in line with earlier observations \citep{ding2021_iqa} – is the importance of using a perceptual loss function for training image-generation models. The top five models were all trained to minimize the perceptual loss and outperformed models trained with either MSE or MAE in the original pixel space.

\paragraph{Advanced Perceptual Metrics with Expert Judgement.}

The five final columns of \tableref{tab:models} reveal that recently proposed advanced perceptual metrics – in particular, DISTS and LPIPS – select the best models, as defined by expert judgement. \tableref{tab:models} reports the metric values on our held-out validation set and we also show that performance is roughly consistent on the official da Vinci test set in \appendixref{test-perf}. Given the high costs associated with obtaining surgeon feedback, this is a positive finding. Future research can rely on DISTS and/or LPIPS for faster and cheaper iteration in designing approaches to 2D-to-3D surgical video reconstruction. 

Unlike DISTS and LPIPS, MSE-based PSNR did not show any discriminative capability among these models, resulting in more or less similar scores for all the tested models. On the other hand, SSIM, which has been successful with natural images (not surgical images), ended up choosing the two worst models, suggesting major differences between natural and surgical images. FID preferred models trained with the perceptual loss, perhaps unsurprisingly because FID itself is a perceptual loss. However, FID ended up favoring the pixel-shift baselines over any of the learned models, which significantly limits the applicability and reliability of FID. See \appendixref{baseline} for more details.

These observations are further confirmed by the rank correlations among automatic metrics and experts, as shown in \tableref{tab:spear_rank_corr}. The rankings of the models by both DISTS and LPIPS correlate
almost perfectly with the expert ranking. This correlation does not exist with the other automatic metrics tested. 

\begin{table}[htbp]
\setlength{\tabcolsep}{3pt}
\floatconts
    {tab:spear_rank_corr}
    {\caption{Spearman's rank correlation coefficient between expert and automatic metric rankings.}}
    {\begin{tabular}{lcccccc}
    \toprule 
        & Expert & DISTS & LPIPS & FID & SSIM & PSNR  \\
        \midrule
        Expert  & 1.0   & \textbf{0.98} & \textbf{0.95} & 0.76 & 0.64 & 0.41 \\ 
        DISTS   &       & 1.0           & 0.98          & 0.71 & 0.60 & 0.39 \\ 
        LPIPS   &       &               & 1.0           & 0.74 & 0.67 & 0.52 \\ 
        FID     &       &               &               & 1.0  & 0.93 & 0.73 \\
        SSIM    &       &               &               &      & 1.0. & 0.88 \\ 
        PSNR    &       &               &               &      &      & 1.0  \\ 
        \bottomrule
    \end{tabular}}
\end{table}

\paragraph{Qualitative Inspection.}

In order to get a better sense of how models differ, we manually inspect selected output frames. \figureref{fig:qual_rank_comparison} shows a representative example, displaying cropped tiles from the outputs of the eight ranked models. These are ordered according to their ranks (best to worst). Full frames and additional examples are found in \appendixref{sec:image_survey_more}.

The models trained using the perceptual loss produce visibly sharper images. One can easily discern visual details in the gauze in examples (b--f) that are not present in examples (g--i), the latter of which were produced by the only three models that did not use the perceptual loss. 

\begin{figure*}[htbp]
\floatconts
    {fig:qual_rank_comparison}
    {\caption{Cropped regions from frames generated by the different models based on the same input. \protect\subfigref{fig:target} Target right view. \protect\subfigref{fig:1}-\protect\subfigref{fig:8} Model results ordered by ranking, from left to right (best to worst).}}
    {%
    \subfigure[][b]{\label{fig:target}%
        \includegraphics[width=0.18\linewidth]{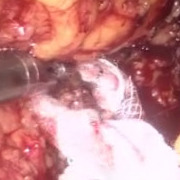}}
    \subfigure[][b]{\label{fig:1}%
        \includegraphics[width=0.18\linewidth]{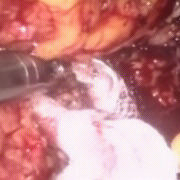}}
    \subfigure[][b]{\label{fig:2}%
        \includegraphics[width=0.18\linewidth]{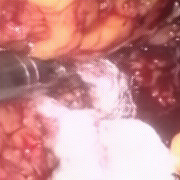}}
    \subfigure[][b]{\label{fig:3}%
        \includegraphics[width=0.18\linewidth]{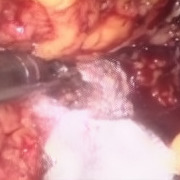}}
    \subfigure[][b]{\label{fig:4}%
        \includegraphics[width=0.18\linewidth]{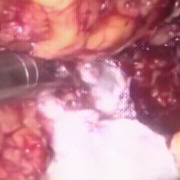}}
    \hspace*{.167\textwidth}\quad
    \subfigure[][b]{\label{fig:5}%
        \includegraphics[width=0.18\linewidth]{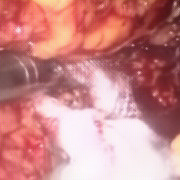}}
    \subfigure[][b]{\label{fig:6}%
        \includegraphics[width=0.18\linewidth]{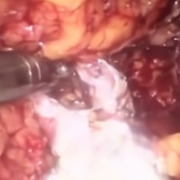}}
    \subfigure[][b]{\label{fig:7}%
        \includegraphics[width=0.18\linewidth]{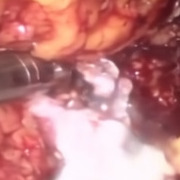}}
    \subfigure[][b]{\label{fig:8}%
        \includegraphics[width=0.18\linewidth]{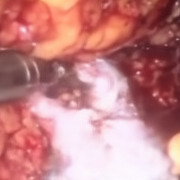}}
    }
\end{figure*}

$$
$$

\paragraph{Expert Judgement vs. Non-Expert Judgement.}

As additional analysis, we conducted another frame-level reader study with 10 non-expert readers with no medical expertise. We calculate how often the majority vote of the experts agrees with the majority vote of the non-experts over all questions. We found that, for 80\% of the questions, the expert and non-expert majority votes were in agreement. 
We attribute this high level of agreement to visual anchors common for both the expert and non-expert readers, such as the legibility of text or the sharp edges of a surgical tool. Disagreements between experts and non-experts, on the other hand, seem to be caused by certain features that are visually subtle but that impact surgical outcomes and whose presence surgeons are therefore trained to search for. For instance, two surgeons commented on the appearance of the vasculature in the generated images. Although vasculature is critical in surgical processes, since blood vessels bleed when cut, non-experts are unlikely to notice it in a given image.

The average within-group disagreement was slightly higher among the non-experts (0.26) than among the experts (0.17), where the within-group disagreement was calculated as the percentage of members who disagreed with the majority vote of the group, averaged over all questions. We conjecture that this lower within-group disagreement rate among the experts is also due to their attention to surgically relevant details that are not taken into account by non-expert readers. 

Based on this comparison between the expert and non-expert readers, we conclude that one cannot fully substitute expert judgment with non-expert judgement. Nevertheless, reasonably high agreement indicates that non-expert surveys – which are easier to come by than expert surveys – can be used to iterate more efficiently in future work.

\subsection{VR Video Reader Study}

We choose the top three models from \tableref{tab:models} in addition to the best MSE model and the best MAE model for testing in the second reader study, a VR video survey. We recruit two surgeons for this study. The full version of \tableref{tab:vr_results} includes detailed individual surgeon comments and can be found in \appendixref{vr_vid_supp}.

\begin{table}[h]
\floatconts
    {tab:vr_results}
    {\caption{Expert evaluation in VR. Scores range from 1 (worst) to 5 (best). }}
    {\begin{tabular}{lcc}
    \toprule
    Model description & Avg & StdDev \\
    \toprule
    5 fr, perceptual, bilinear  & 4.67 & 0.52 \\
    \midrule
    5 fr, perceptual            & 3.83 & 0.98 \\
    \midrule
    5 fr, perceptual, 6 layers  & 3.50 & 0.55 \\
    \midrule
    1 fr, MSE, 6 layers         & 2.83 & 0.75 \\
    \midrule
    1 fr, MAE                   & 2.83 & 0.98 \\
    \bottomrule
    \end{tabular}}
\end{table}

The first finding from this reader study is that both surgeons strongly favored the 3D reconstructed surgical video clips over the corresponding 2D clips. This supports the validity and importance of the proposed task of 2D-to-3D reconstruction in surgical video and demonstrates the effectiveness of the proposed U-Net–based approach to this problem. 

The second finding is that models trained with the perceptual loss were strongly preferred over models trained with MSE or MAE, which confirms our observation from the first reader study. Comments by the readers suggest that one major cause of this distinction is the blurriness associated with pixel-wise losses such as MSE or MAE. In particular, two comments describing the MSE-trained model mention the lack of detail in the right periphery. 

We observe that the favorite model among the experts in the second reader study was the third model from the first, frame-level reader study. The major difference between this model and the other two models, also trained with the perceptual loss, is that it uses bilinear interpolation for upsampling in the decoder rather than transposed convolution. In order to study the difference between these two upsampling implementations, we manually inspect the generated frames from the models (one with bilinear upsampling and the other with transposed convolution). Doing so revealed that transposed convolution tends to amplify checkerboard artifacts, as shown in \figureref{fig:bilinear}, which is in line with earlier observations by \citet{odena2016deconvolution}. Although this effect is subtle at the frame level, we suspect that, when viewed in 3D, the artifacts produce a more disorientating visual experience.

\begin{figure}[htbp]
\floatconts
    {fig:bilinear}
    {\caption{Visual artifacts associated with upsampling layers. All models use five frames, a 3D convolutional temporal module, and perceptual loss. Models differ only in the upsampling layers. (a) and (b) use transposed convolution (TC), with six and five layers respectively, while (c) has five layers of bilinear interpolation (BI).}}
    {%
    \subfigure[TC (6)][b]{\label{fig:u22}%
        \includegraphics[width=0.25\linewidth]{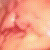}}
    \quad
    \subfigure[TC (5)][b]{\label{fig:u3}%
         \includegraphics[width=0.25\linewidth]{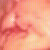}}
    \quad
    \subfigure[BI (5)][b]{\label{fig:u18}%
         \includegraphics[width=0.25\linewidth]{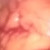}}
    }
\end{figure}

\section{Conclusion}
Minimally invasive endoscopic surgery is one of the most important surgical advances of the past half-century. The use of endoscopy, however, is often limited by the steep learning curve associated with a lack of depth perception, and the challenge of delivering high-quality visualization via increasingly smaller optical instruments. Enhanced endoscopic imaging through computer vision has the potential to further democratize minimally invasive surgery.

In this paper, we proposed the task of converting 2D endoscopy video to 3D video and investigated the feasibility of deep learning for tackling the lack of depth perception in surgical endoscopy. More specifically, we designed modified U-Nets to generate a novel perspective (right view) given a series of consecutive left-view frames and tested their effectiveness by running an extensive set of reader studies, both frame-level and video-level, with experienced surgeons.

The reader studies along with thorough analysis revealed a few major findings. First, surgeons preferred the generated stereo video over the corresponding 2D video. This confirms the usefulness of the proposed task and the effectiveness of the proposed deep learning–based solution. Second, the models that were favored by expert surgeons were all trained with a perceptual loss and utilized multiple consecutive frames of the observed view with a convolutional temporal module. This finding 1) verifies our hypothesis that temporal information, available from multiple past frames, is critical in reconstructing the missing view in the stereo pair, and 2) shows that the proposed variant of the U-Net is able to exploit such information from input frames. Third, we identified two perceptual loss functions, DISTS and LPIPS, that correlate strongly with expert judgement. These automatic metrics will enable rapid iteration to improve algorithms for solving the proposed task without relying on time-consuming and costly reader studies. Finally, we demonstrated the overall importance of expert readers compared to non-expert readers (without any medical experience) when assessing 2D-to-3D reconstruction algorithms for surgical video. Overall, these findings indicate that the proposed task is feasible and useful, and that the proposed approach, despite being the very first attempt at this problem, is a promising one.

\paragraph{Future Directions.}

The positive findings from this paper present us with a number of future directions. First, running another set of video-based reader studies with higher-resolution reconstruction and higher-quality VR gear would further strengthen the usefulness of both the proposed task and the proposed approach. Second, better network architectures as well as learning hyperparameters should be sought out, now that our experiments have revealed advanced perceptual loss to be a good proxy for expensive human evaluation. Third, we expect more advanced temporal modules, such as a recurrent DNN-based module, to generate better images by capturing longer-range dependencies. Finally, this task is not necessarily limited to endoscopy, and we anticipate that our findings will be transferable to similar problems in medicine and other domains. 

\bibliography{main}

\appendix

\section{Hamlyn da Vinci Dataset} \label{data}
The {\it da Vinci} dataset from the Hamlyn Center for Robotic Surgery \citep{ye2017selfsupervised} contains rectified stereo images of size $384 \times 192$ pixels partitioned into 34,241 train and 7,192 test image sequences from video. To generate train and validation data splits, the full train image sequence is split into sequences of 1000 images. After this, 80\% of the splits are randomly allocated to the train set and the remaining 20\% are allocated to the validation set. To generate samples from each data split, a sliding window of length $K$, where $K$ is the image input sequence length for the model, is applied to each sequence of 1000 images in their respective data split. Individual samples are shuffled again before training.

The dataset does not disclose the frame rate of the original video stream. We found that generating video from the dataset images at 20 frames per second yielded realistic results for evaluation.

\begin{figure}[htbp]
\floatconts
    {fig:full_ref}
    {\caption{A sample of the left and right views from the dataset released by the Hamlyn Center for Robotic Surgery \citep{ye2017selfsupervised}.}}
    {
     \subfigure[Left view (input)][]{\label{fig:left}
         \includegraphics[width=0.45\textwidth]{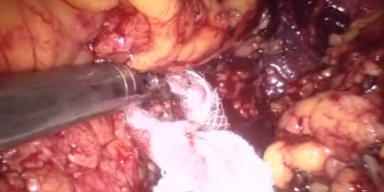}}
     \subfigure[Right view (target)][]{\label{fig:right}
         \includegraphics[width=0.45\textwidth]{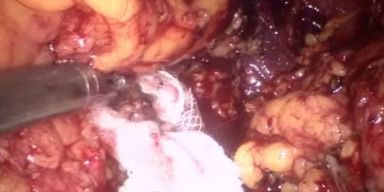}}
    }     
\end{figure}

\section{Test Set Performance} \label{test-perf}

We verify that the model performance in terms of DISTS and LPIPS, the two most successful automatic metrics, is consistent on the official Da Vinci test dataset.

\begin{table}[htbp]
\setlength{\tabcolsep}{3pt}
\floatconts
    {tab:test_set_perf}
    {\caption{Perceptual metric performance on official da Vinci test set.}}
    {\begin{tabular}{llcc}
    \toprule 
        Rank & Model & DISTS & LPIPS \\
        \midrule
        1  & 5 fr+perceptual+6 layers     & 0.129 & 0.148 \\ 
        2  & 5 fr+perceptual              & 0.141 & 0.155 \\ 
        3  & 5 fr+perceptual+bilinear     & 0.147 & 0.166 \\ 
        4  & 1 fr+perceptual              & 0.156 & 0.180 \\
        5  & 10 fr+perceptual             & 0.142 & 0.167 \\ 
        6  & 1 fr+MSE+6 layers            & 0.157 & 0.194 \\ 
        7  & 1 fr+MAE                     & 0.159 & 0.200 \\
        8  & 5 fr+MSE+6 layers            & 0.152 & 0.187 \\
        \bottomrule
    \end{tabular}}
\end{table}

\section{Baseline Results} \label{baseline}
Baseline results were generated using the procedure described in \sectionref{sec:baseline_setup}. Although most automatic metrics ranked baseline models poorly when compared to U-Net based models, FID failed as a reliable measure in this study. The FID metric ranked the baseline models higher than the trained models (\tableref{tab:baselinemodels}). Images generated by the baseline models (\figureref{fig:baseline_imgs}) contain discontinuities, a duplicated strip of the original image or a black strip in place of the missing pixels, and a constant global disparity. Since the baseline models are not candidate solutions to the task of 2D-to-3D reconstruction, we find FID using 2048 final average pooling features to be unsuitable in the context of our task.

\begin{table*}[htbp]
\floatconts
    {tab:baselinemodels}
    {\caption{Constant global disparity baseline results optimized on the validation set. Results are reported on the validation set.}}
    {
    \begin{tabular}{llccccc}
    \toprule 
        Baseline Model & Pixel Shift & DISTS & LPIPS & FID & SSIM & PSNR  \\
        \midrule
        Copy pixels, perceptual loss    & -31   & 0.129 & 0.201 & 30.079  & 0.493 & 17.675 \\ 
        Copy pixels, MSE loss           & -37   & 0.128 & 0.205 & 32.648  & 0.486 & 17.728 \\ 
        Copy pixels, MAE loss           & -37   & 0.128 & 0.205 & 32.648  & 0.486 & 17.728 \\ 
        Fill zeros, perceptual loss     & -384  & 0.800 & 0.840 & 549.658 & 0.002 & 5.254  \\
        Fill zeros, MSE loss            & -32   & 0.145 & 0.237 & 35.148  & 0.458 & 16.067 \\ 
        Fill zeros, MAE loss            & -33   & 0.146 & 0.237 & 35.655  & 0.455 & 16.043 \\ 
        \bottomrule
    \end{tabular}
    }
\end{table*}

\begin{figure}[h]
\floatconts
    {fig:baseline_imgs}
    {\caption{Examples of baseline outputs with global disparity and basic in-painting methods.}}
    {
     \subfigure[Left view (input)][]{\label{fig:baseline_left}
         \includegraphics[width=0.35\textwidth]{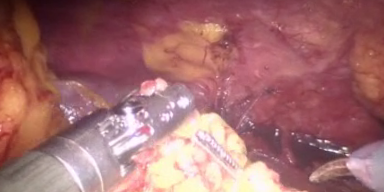}}
     \qquad
     \subfigure[Right view (target)][]{\label{fig:baseline_right}
         \includegraphics[width=0.35\textwidth]{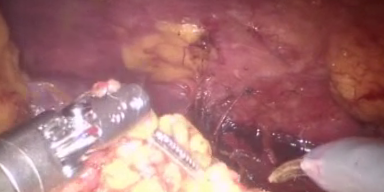}}

     \subfigure[Baseline with original pixel fill][]{\label{fig:baseline_shift_fill}
         \includegraphics[width=0.35\textwidth]{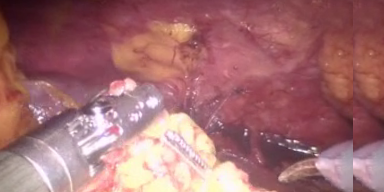}}
    \qquad
    \subfigure[Baseline with black fill][]{\label{fig:baseline_black_fill}
        \includegraphics[width=0.35\textwidth]{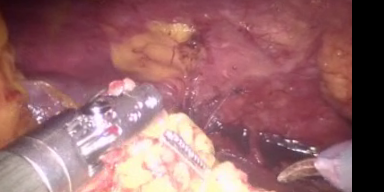}}

    } 
\end{figure}

\section{Image Reader Study}
\subsection{Initial model screening for reader study} \label{sec:app_screening}

Eight models (\tableref{tab:selected_models}) were selected for the frame-level reader study from an initial set of over 40 different configurations with varying architecture and loss as presented in \tableref{tab:configs}. As part of the effort to identify automatic metrics for model selection, we selected the top eight models based on four metrics: LPIPS, FID, SSIM and PSNR. In addition, we ran multiple multiple rounds of qualitative assessments with the supervising surgeon to assess perceptual differences and rankings of the generated images as additional screening for the reader study.

\begin{table*}[ht]
\floatconts
    {tab:configs}
    {\caption{A summary of the U-Net architecture and losses tested in the initial search for candidate models for the expert reader study.}}
    {
    \begin{tabular}{ll}
    \toprule 
    \bfseries Model Component & \bfseries Tested Configurations \\
    \midrule
        Loss function & MSE, MAE, Perceptual \\
        Temporal structure & channel-stacking, spatio-temporal, 3D convolutions, max-pooling, avg-pooling \\
        Number of input frames & 1, 5, or 10 \\
        U-Net depth & 5 or 6 blocks \\
        Upsampling mechanism & transposed convolution or bilinear interpolation upsampling \\
        Sigmoid output & Yes/No \\
        Extra skip connection & Yes/No \\
        \bottomrule
    \end{tabular}
    }
\end{table*}

\begin{table}[h]
\setlength{\tabcolsep}{3pt}
\floatconts
  {tab:selected_models}%
  {\caption{The eight U-Net models selected for the frame-level reader study. Unless specified otherwise, each model used five double convolution layers and used transpose convolution for upsampling. If multiple input frames were used, a spatio-temporal U-Net architecture was used with 3D convolution temporal module.}}%
  {\begin{tabular}{lll}
    \toprule 
        \bfseries Model description & \bfseries Loss \\
        \midrule
        1 frame & Perceptual \\
        1 frame & MAE \\
        1 frame, 6 layers & MSE \\
        5 frames & Perceptual \\
        5 frames, 6 layers & Perceptual \\
        5 frames, 6 layers & MSE \\
        5 frames, bilinear upsampling & Perceptual \\
        10 frames & MSE \\
    \bottomrule
  \end{tabular}}
\end{table}

All spatio-temporal models based on maximum or average pooling temporal modules were eliminate from the study as the generated images were poor quality (\figureref{fig:max_avg_pool}). Stacking multiple frames along the channel axis did not result in significant qualitative or quantitative improvements and were also filtered from the study. Thus, all multi-frame models included in the reader study used a spatio-temporal architecture with 3D convolutions.

\begin{figure}[p]
\floatconts
  {fig:max_avg_pool}
  {\caption{Models using maximum \protect\subfigref{fig:perc_avg} or average pooling  \protect\subfigref{fig:perc_avg} temporal modules were found to be unsuitable for the task due to poor quantitative and qualitative performance when compared to the target output \protect\subfigref{fig:target_1}}.
  }
  {%
    \subfigure[][]{\label{fig:target_1}
        \includegraphics[width=0.45\linewidth]{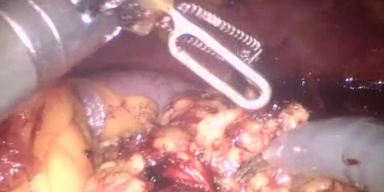}}
    \qquad
    \subfigure[][]{\label{fig:perc_avg}%
      \includegraphics[width=0.45\linewidth]{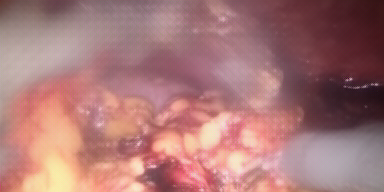}}%
    \qquad
    \subfigure[][]{\label{fig:perc_max}%
      \includegraphics[width=0.45\linewidth]{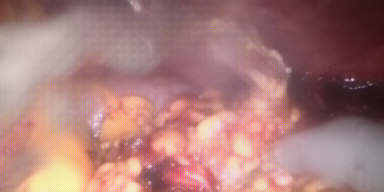}}
  }
\end{figure}

\subsection{Image-based reader study format} \label{sec:image_survey_more}
\begin{figure}[htbp]
    \centering
    \includegraphics[width=0.95\textwidth]{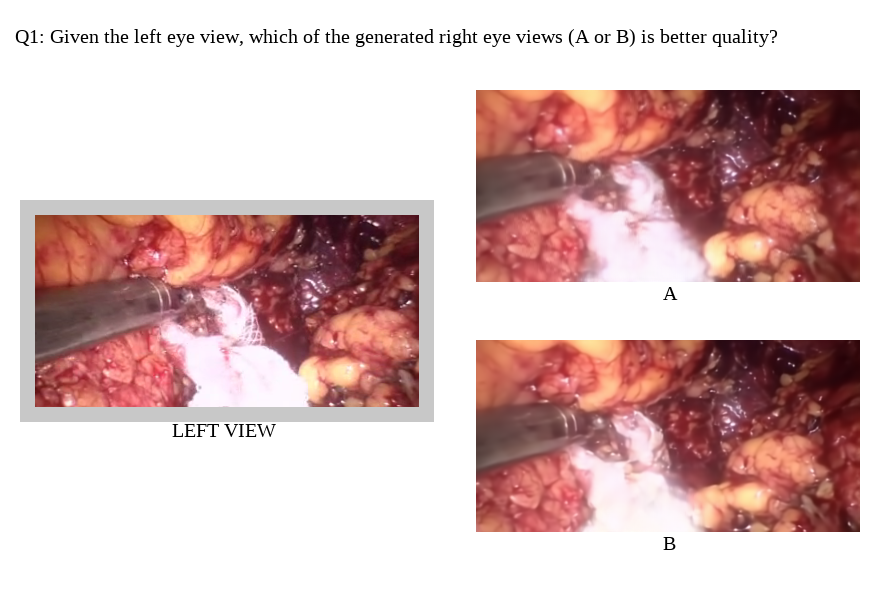}
    \caption{Reader study participants were shown a series of two-alternative forced choice questions with no time limit and asked to select the higher quality candidate stereo image with the question: \textit{Given the left eye view, which of the generated right eye views (A or B) is better quality?}}
    \label{fig:image_survey}
\end{figure}

The image-based reader study was completed remotely by subjects using a Google Forms survey. The first page summarized the background, goals, and methods for the project. In addition, we state instructions for the survey. The second page of the survey requests the subject's profession and if they have performed endoscopic surgery (2D and/or 3D). On the third page, readers are presented a set of instructions, a set of questions, and a link to the directory containing the image comparison questions (\figureref{fig:image_survey}).

\subsection{Qualitative Examples}
\begin{figure}[ht]
\floatconts
    {fig:top_eight_qual}
    {\caption{Generated images from the top eight models as ranked by the image-based reader study. Models are ranked from best \protect\subfigref{fig:111} to worst \protect\subfigref{fig:888}. Refer to \tableref{tab:models} for model descriptions and model performance.}}
    {
    \subfigure[][]{\label{fig:111}
        \includegraphics[width=0.4\textwidth]{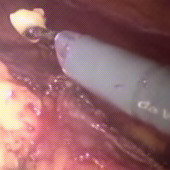}}
    \qquad
    \subfigure[][]{\label{fig:222}
        \includegraphics[width=0.4\textwidth]{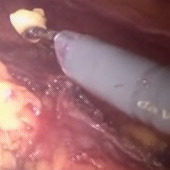}}

    \subfigure[][]{\label{fig:333}
        \includegraphics[width=0.4\textwidth]{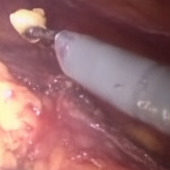}}
    \qquad
    \subfigure[][]{\label{fig:444}
        \includegraphics[width=0.4\textwidth]{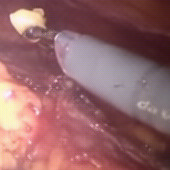}}

    \subfigure[][]{\label{fig:555}
        \includegraphics[width=0.4\textwidth]{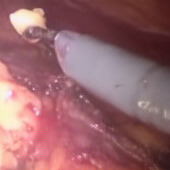}}
    \qquad
    \subfigure[][]{\label{fig:666}
        \includegraphics[width=0.4\textwidth]{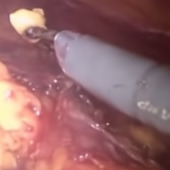}}

    \subfigure[][]{\label{fig:777}
        \includegraphics[width=0.4\textwidth]{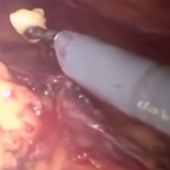}}
    \qquad
    \subfigure[][]{\label{fig:888}
        \includegraphics[width=0.4\textwidth]{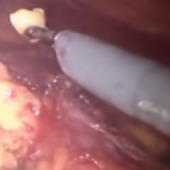}}
    }
\end{figure}


\subsection{Multi-frame models vs. single frame models} \label{sec:multi_vs_single}
To validate that increased performance for multi-frame models was not due to increased model complexity (3$\times$ increased model complexity when compared to the equivalent single frame model), we re-train the best multi-frame model (as ranked by LPIPS and DISTS and the reader study) with the repeated frames as input. This repeated frame model resulted in significantly lower LPIPS and DISTS values than the equivalent model conditioned on past frames.

\subsection{Bradley Terry ranking model}
\label{sec:bradleyterry}
Given a set of models to be ranked, the probability of selecting the output from some model $i$ from a presented pair of outputs from model $i$ and $j$ is \[P(i) = \frac{\alpha_i}{\alpha_i + \alpha_j}\] where $\alpha_i$ is the \textit{worth} of item $i$. 
We use the PlackettLuce package \citep{PlackettLuce20}, which reduces to the Bradley-Terry model given a list of pairwise-comparisons, to estimate relative rankings of the reader study models. 
We estimate worth and bounded quasi-standard errors \citep{firth2003} for each model and transform them to log-scale. Since we expect all models compared to have worth greater than 0, we transform worth to the log scale to provide bounds on confidence intervals such that then are constrained to be greater than 0. 
We choose the target images as the reference such that the quasi-standard errors for the other models can be estimated.

\section{VR Video Reader Study}\label{vr_vid_supp}

\begin{table*}[ht!]
\setlength{\hfuzz}{1.1\columnwidth}
\begin{minipage}{\textwidth}
\floatconts
    {tab:vr_expanded_results}
    {\caption{We expand \tableref{tab:vr_results} and include informative comment excerpts from the readers for VR model evaluation. For all models, both surgeons preferred the generated stereo video compared to the raw 2D alternative.}}
    {\begin{tabular}{lccp{8cm}}
    \toprule
    Model description & Avg & StdDev & Comments \\
    \toprule
    5 fr, perceptual, bilinear  & 4.67 & 0.52 & \parbox{8cm}{``Perfect.'' \\ ``Great.''} \\
    \midrule
    5 fr, perceptual            & 3.83 & 0.98 & \parbox{8cm}{``Really like this one.'' \\ ``Temporal inconsistencies in grasper with shearing between frames''} \\
    \midrule
    5 fr, perceptual, 6 layers  & 3.50 & 0.55 & \parbox{8cm}{``Loves it in 3D. Surprised at how much I like it.''\\
    ``Looked good. Couldn't see text on instrument.''}\\
    \midrule
    1 fr, MSE, 6 layers         & 2.83 & 0.75 & \parbox{8cm}{``Periphery is not as crisp.''\\``Right periphery is not very good.''\\``Blurrier, little disorienting.''\\``Globally less sharpness. Less details. Less pronounced stereoscopic effect''} \\
    \midrule
    1 fr, MAE                    & 2.83 & 0.98 & \parbox{8cm}{``Depth perception was not as good as the first round.''\\``Blurrier, still pretty good.''\\``Not as detailed.''} \\
    \bottomrule
    \end{tabular}}
\end{minipage}
\end{table*}


We selected the three top ranked models from \tableref{tab:models} as well as the top MAE and MSE models to generate stereo video for the VR video reader study. We selected a diverse set of models to determine if image-based reader study results match video-based reader study results. 

In the VR video reader study, subjects were given a Google Cardboard headset with an iPhone 12 Pro as the video source. Subjects were shown six videos of three scenes of approximately 20 seconds in length per model. Different scenes were selected based on variation in depth, movement, and color in each scene. Subjects were first shown the scene in 2D in which both eyes are shown the exact same video. Subjects are then shown a generated stereo 3D video of the same scene seen previously. Subjects are asked three questions: 1) Would you prefer the first video or second video when doing surgery? 2) Rate the stereo 3D video quality from 1-5 with 5 being best? 3) Provide general feedback about the quality of the video based on your experience in surgery.





\end{document}